\documentclass[aps,pra,amssymb,reprint,twocolumn,groupaddress]{revtex4-1}

\usepackage[ansinew]{inputenc}
\usepackage{graphicx}
\usepackage[breaklinks=false,colorlinks=true,linkcolor=blue,urlcolor=blue,citecolor=blue]{hyperref}
\usepackage{setspace}
\usepackage[ansinew]{inputenc}
\usepackage{color}
\usepackage{caption}
\usepackage[labelfont=bf]{caption}
\usepackage{graphicx}
\usepackage{amsmath}
\usepackage{color}

\begin{document}

\title{Extreme-ultraviolet refractive optics} 

\author{Lorenz Drescher}

\author{Oleg Kornilov}

\author{Tobias Witting}

\author{Geert Reitsma}

\author{Nils Monserud}

\author{Arnaud Rouz\'ee}

\author{Jochen Mikosch}

\author{Marc J. J. Vrakking}

\author{Bernd Sch\"utte}

\affiliation{Max-Born-Institut, Max-Born-Strasse 2A, 12489 Berlin, Germany}


\begin{abstract}
Refraction is a well-known optical phenomenon that alters the direction of light waves propagating through matter. Microscopes, lenses and prisms based on refraction are indispensable tools for controlling the properties of light beams at visible, infrared, ultraviolet and X-ray wavelengths. The large absorption of extreme-ultraviolet (XUV) radiation in matter, however, hinders the development of refractive lenses and prisms in this spectral region. Here, we demonstrate control over the refraction of XUV radiation by using a gas jet with a density gradient across the XUV beam profile. A gas phase prism is demonstrated that leads to a frequency-dependent deflection of the XUV beam. The strong deflection in the vicinity of atomic resonances is further used to develop a deformable XUV refractive lens, with low absorption and a focal length that can be tuned by varying the gas pressure. Our results provide novel opportunities in XUV science and open a route towards the transfer of refraction-based techniques including microscopy and nanofocusing, which are well established in other spectral regions, to the XUV domain.
\end{abstract}

\maketitle

\section*{Introduction}
\noindent Refraction of light is omnipresent in nature, where it forms the basis for the functionality of the human eye and the observation of a rainbow. Refraction is also exploited in many applications in the visible, infrared and ultraviolet spectral regions. For instance, refractive errors of the eye are corrected by glasses or contact lenses, while optical microscopes enable the magnification of small objects or structures. In the context of laser physics, refractive lenses are extensively used to focus or (de-)magnify laser beams. Dispersion and deflection of light by optical prisms is used to compress or stretch ultrashort laser pulses.

When R\"ontgen discovered X-rays in 1895, he attempted refraction experiments using prisms and lenses~\cite{rontgen95, rontgen96}. Since he observed no significant deflection of the X-rays, he concluded that refractive lenses are not suitable for focusing X-ray radiation. One century later, a compound refractive lens (CRL) consisting of a lens array was nevertheless developed for the hard X-ray regime~\cite{snigirev96, lengeler99}, facilitated by the comparably low absorption in this spectral region. CRLs are used to focus X-rays emitted from modern synchrotron~\cite{elleaume98, santoro14} and free-electron laser facilities~\cite{chollet15, heimann16, tono17, nakatsutsumi14}. They have e.g. been applied for hard X-ray microscopy~\cite{lengeler99b, simons15}, for X-ray nanofocusing~\cite{schroer05}, for the investigation of crystal scattering~\cite{meijer17}, as well as for coherent diffractive imaging of nanoscale samples~\cite{schroer08}. 

In the past decades, a large variety of extreme-ultraviolet (XUV) and soft X-ray sources have been developed both in laboratory environments~\cite{mcpherson87, ferray88, krausz09, rocca99, giulietti98, ditmire95} and at large-scale facilities~\cite{marr13, ackermann07, shintake08, allaria12}. Focusing of XUV and soft X-ray pulses is typically achieved by reflective mirrors and diffractive Fresnel zone plates~\cite{baez60, baez61}. Refractive elements are so far missing in the XUV range, but are highly desirable. Refractive lenses could e.g. be used to focus XUV pulses without changing the XUV propagation direction, thereby providing a high degree of flexibility and facilitating their use. The lack of refractive elements in the XUV spectral region can be explained by the large absorption of radiation in the XUV domain by commonly used materials. The use of specially designed microscopic refractive lenses has been proposed~\cite{wang03, pan16}. However, the need to use very thin lenses with a sophisticated design makes a practical implementation extremely challenging, and no demonstrations exist to date.

In this article we demonstrate that control over the refraction of XUV pulses can be achieved by using gases instead of solids. In a first set of experiments, broadband high-harmonic generation (HHG) pulses are spectrally dispersed and deflected when propagating through a gas jet with a density gradient, which induces an optical path gradient across the XUV beam profile. The operation principle thus resembles that of a prism. In a second set of experiments, we demonstrate the operation of a deformable XUV refractive lens using a cylindrical gas jet. Individual harmonic orders are focused to a spot size that is controlled by the type of gas and the gas pressure. We discuss the potential for the development and application of XUV refractive lenses in the future.


\section*{Gas phase XUV prism}

\begin{figure*}[tb]
 \centering                                          
  \includegraphics[width=\textwidth]{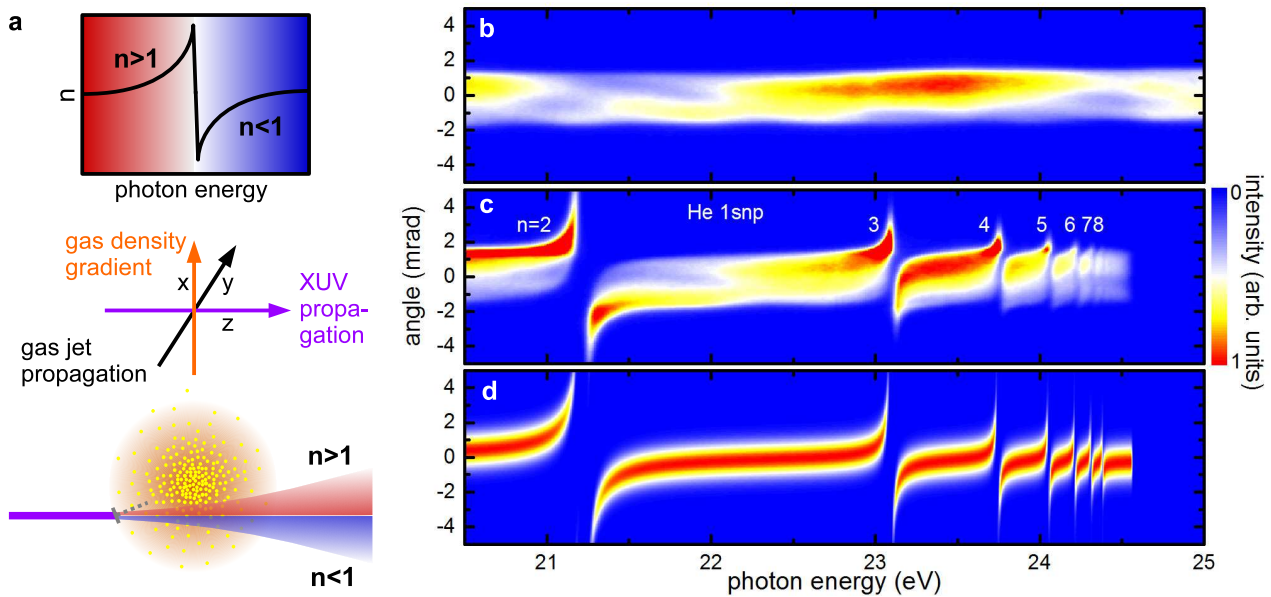}
 \caption{\label{figure1} \textbf{XUV refractive prism.} \textbf{a}, Top panel: Dispersive lineshape of the refractive index in the vicinity of an atomic resonance. Spectral components at photon energies below the resonance ($n>1$) are referred to by a red color, whereas spectral components at photon energies above the resonance ($n<1$) are referred to by a blue color. Middle panel: Experimental configuration, showing an XUV pulse (violet arrow) that crosses a gas jet (black arrow), which has a density gradient in the vertical direction (orange arrow), at right angles. Bottom panel: Deflection of an XUV pulse propagating below the center of the gas jet. \textbf{b}, Angle-resolved spectrum of a broadband HHG pulse measured in the absence of the gas jet. The angular divergence of the XUV beam in the vertical direction is reflected in the spatial distribution along the vertical axis. \textbf{c}, The same spectrum after propagation at a distance of 0.3~mm below the center of a dense He gas jet (generated using a backing pressure of 10~bar) shows clear signatures of refraction. Spectral components with photon energies just below the 1s$n$p resonances of He are deflected upwards, whereas spectral components just above these resonances are deflected downwards. The deflection angles are largest close to the 1s2p resonance and decrease for higher resonances due to the decreasing oscillator strengths. Above the ionization potential of He (at 24.58~eV), the XUV radiation is strongly absorbed. \textbf{d} Simulation of the XUV refraction in an inhomogeneous He gas jet, taking into account 1s$n$p resonances with $n=2,3 \ldots 8$. The simulation indicates that using a backing pressure of 10~bar, a gas jet with a peak density of $9\times 10^{19}$~atoms/cm$^3$ (corresponding to a pressure of 3.7~bar at 300~K) was achieved in the interaction zone.}
\end{figure*}


\noindent In the vicinity of atomic resonances, the refractive index $n$ exhibits a dispersive lineshape as depicted in the top part of Fig.~\ref{figure1}a. As the photon energy approaches the resonant energy, $n$ first increases, and then steeply decreases across the resonance to values below unity, before increasing again~\cite{leroux62, roschdestwensky12, korff32}.


Our scheme for control over the refraction of XUV pulses employing an inhomogeneous gas target is presented in the middle and bottom panels of Fig.~\ref{figure1}a. The XUV pulses pass through a gas jet, which propagates in a direction perpendicular to the XUV beam and which has a density gradient in the vertical direction (middle panel in Fig.~\ref{figure1}a). When the XUV pulse crosses the gas jet off-centre, the latter acts as a prism and induces angular dispersion and deflection of the XUV radiation. For an XUV beam that is incident below the center of the gas jet, spectral components of the beam for which $n>1$ are deflected upwards (red color in the bottom panel of Fig.~\ref{figure1}a), whereas spectral components of the beam, for which $n<1$, are deflected downwards (blue color in the bottom panel of Fig.~\ref{figure1}a). 


An experimental demonstration of this concept is presented in Fig.~\ref{figure1}b,c. Fig.~\ref{figure1}b shows an HHG spectrum produced using near-infrared (NIR) pulses with a duration of 4.5~fs, as measured on a 2D detector, where the horizontal axis represents the axis along which the XUV spectrum is dispersed using a flat-field grating spectrometer (see Methods). When the broadband HHG pulses propagate 0.3~mm below the center of a dense He gas jet, the XUV spectrum is strongly modified, see Fig.~\ref{figure1}c. Spectral components below the 1s$n$p ($n=2,3,\dots$) resonances of He are deflected upwards, whereas spectral components above the 1s$n$p resonances are deflected downwards. 

Microscopically, refraction is explained in terms of oscillating electric dipoles induced by the XUV pulse. The incoming XUV pulse excites atoms that re-emit radiation at the same photon energy (free induction decay)~\cite{bloch46, hahn50, brewer72, wu16, bengtsson17}. Our prism exploits the fact that this re-emitted radiation is phase-shifted with respect to the exciting XUV pulse. By inducing a gas density gradient, the upper part of the XUV pulse acquires a different phase shift than the lower part, which leads to a tilt of the XUV wavefront. The wavefront tilt depends on the refractive index and therefore increases close to a resonance~\cite{liao15}. For example, at a photon energy of 21~eV, i.e. 0.22~eV below the 1s2p resonance, the wavefront tilt is 0.07~degrees at the gas density used in the present experiment. Different spectral components are thus deflected into different directions, with the stronger deflection occurring close to a resonance.


We have simulated refraction in a gas jet using the Lorentz-Lorenz formula (see Methods), assuming an XUV beam with a Gaussian spatial profile and using the properties of the 1s$n$p absorption series of He~\cite{chan91}. The phase accumulated by the XUV pulse during propagation through the gas medium is calculated using the eikonal approximation~\cite{born99}. Propagation of the XUV beam in free space is calculated using the small-angle approximation to the Kirchhoff's diffraction formula solved by the Fourier transform method (see Methods for details). As shown in Fig.~\ref{figure1}d, the simulation reproduces the experimental result well.

\begin{figure}[tb]
 \centering                                          
  \includegraphics[width=8.6cm]{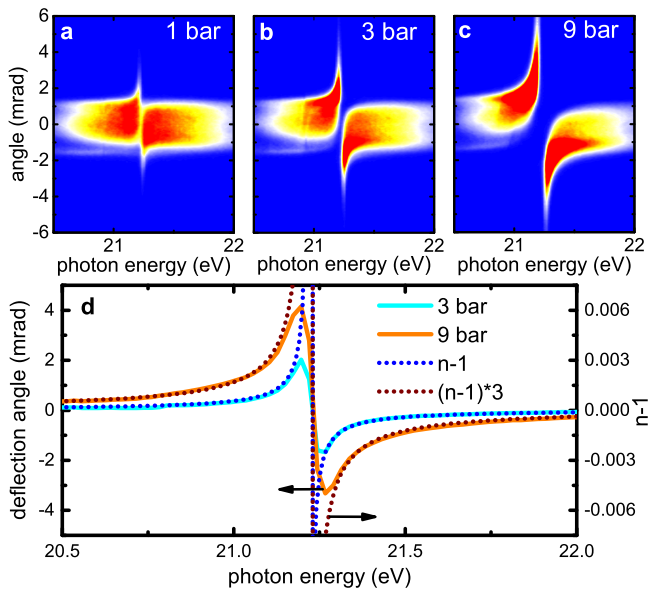}
 \caption{\label{figure2} \textbf{Control over the XUV deflection by the gas pressure.} Angle-resolved XUV spectra after propagation at a distance of 0.3~mm below the center of a He gas jet using backing pressures of \textbf{a}, 1~bar, \textbf{b}, 3~bar and \textbf{c}, 9~bar. \textbf{d}, The average deflection angle as a function of the photon energy for backing pressures of 3~bar (corresponding to a peak pressure in the interaction zone of about 1~bar, cyan solid curve) and 9~bar (orange solid curve). For comparison, the calculated refractivity (i.e. $n-1$) at standard temperature (273.15~K) and standard pressure (1~bar) is plotted on top of the deflection results (blue dotted curve). Note that the calculated refractivity is proportional to the pressure.}
\end{figure}

The deflection of the XUV beam can be controlled by varying the gas pressure. Angle-resolved spectra for He backing pressures of 1~bar, 3~bar and 9~bar are presented in Fig.~\ref{figure2}a-c and show increasing deflection for increasing backing pressure. The average deflection angle as a function of the photon energy (determined by comparing the center-of-mass of the distribution along the vertical axis with and without the He gas jet) is plotted in Fig.~\ref{figure2}d for backing pressures of 3~bar (cyan solid curve) and 9~bar (orange solid curve). For small angles, the deflection is proportional to the refractivity (i.e. $n-1$), which was calculated using the Lorentz-Lorenz formula (see Methods). The shapes of the measured deflection angles (solid curves) and the calculated refractivities (dotted curves) agree well, apart from the region near resonance, where the angular acceptance and the resolution of the XUV spectrometer are no longer sufficient.

\section*{XUV refractive lens}

\begin{figure*}[tb]
 \centering                                          
  \includegraphics[width=18cm]{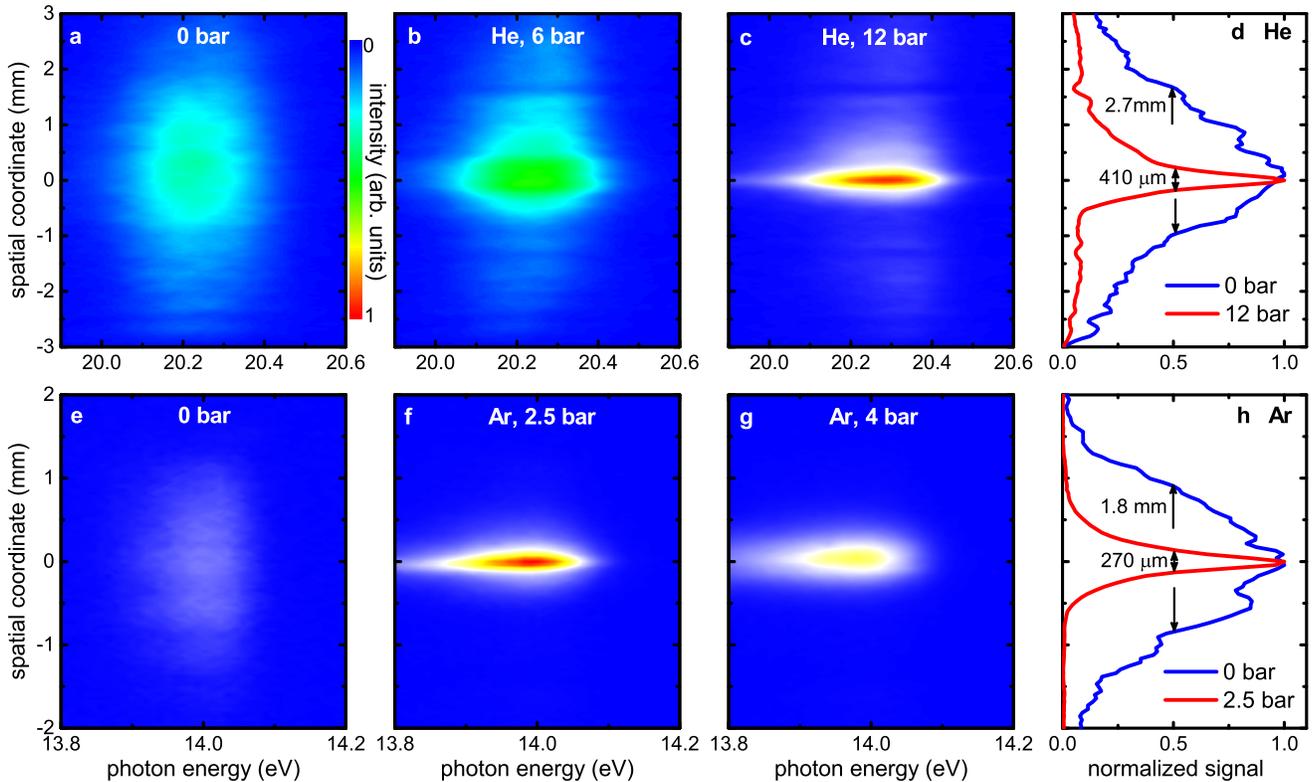}
 \caption{\label{figure3} \textbf{XUV refractive lens.} \textbf{a}, Spatially-resolved spectrum of the unfocused 13th harmonic. The divergence of this harmonic is altered after propagation through a He gas jet, which results in a smaller vertical beam size on the detector, as shown for backing pressures of \textbf{b}, 6~bar and \textbf{c} 12~bar. \textbf{d}, Comparison between the vertical beam profile using backing pressures of 0~bar (blue curve) and 12~bar (red curve). \textbf{e}, Spatially-resolved spectrum of the 9th harmonic, whose photon energy is close to the 3d and 5s resonances of Ar. \textbf{f}, Focusing of this harmonic is achieved by an Ar gas jet at a backing pressure of 2.5~bar. \textbf{g}, When further increasing the backing pressure to 4~bar, an increasing beam size is observed, since the Ar lens focuses the XUV beam between the gas jet and the detector. \textbf{h}, The vertical beam profile of the 9th harmonic for Ar backing pressures of 0~bar (blue curve) and 2.5~bar (red curve). }
\end{figure*}

The deflection of XUV pulses in the vicinity of atomic resonances can be exploited for the design of an XUV refractive lens. In a second set of experiments, high harmonics with a narrow bandwidth of $100-200$~meV and a low beam divergence were generated using a 12~m long beamline~\cite{schutte14a}, see Methods. Fig.~\ref{figure3}a depicts the spatially-resolved spectrum of the 13th harmonic as recorded at a distance of 6~m behind the HHG cell. The photon energy of 20.2~eV is about 1~eV below the 1s2p resonance of He. The spatial extension of the harmonic along the vertical axis (2.7~mm, see Fig.~\ref{figure3}d) corresponds to a full width at half maximum (FWHM) divergence of 0.45 mrad. When a He gas jet with a parabolic profile~\cite{semushin01} and a spatial extension of about 2.5~mm (i.e. similar to the XUV beam diameter, which is 2.3~mm at this point) is placed 0.9~m in front of the detector, the former acts as a lens. Fig.~\ref{figure3}b,c demonstrates focusing of the 13th harmonic for two different backing pressures. Fig.~\ref{figure3}d shows that the FWHM in the vertical direction is reduced from 2.7~mm to 410~$\mu$m by operating the gas jet at a backing pressure of 12~bar (the highest pressure used in the experiment, leading to a peak density in the experiment of about $1\times 10^{20}$~atoms/cm$^3$). We found that absorption of the XUV beam by the He lens is small, i.e. below the estimated detection threshold of 5~$\%$. The geometry of the current experiment leads to focusing in one dimension, analogous to the focusing by a cylindrical lens. A sequence of two perpendicularly placed gas jets, each with a cylindrically shaped density gradient, could be used to focus XUV pulses both horizontally and vertically.


As the deflection of XUV radiation increases for photon energies approaching an atomic resonance, we have studied another example using the 9th harmonic at a photon energy of 14.0~eV (see Fig.~\ref{figure3}e), which is close to the 5s (at 14.09~eV) and 3d (at 14.15~eV) resonances of Ar. In this case, an Ar gas jet with a moderate backing pressure of 2.5~bar (corresponding to a peak density in the interaction region of about $2\times 10^{19}$~atoms/cm$^3$) was used to focus the XUV radiation, as shown in Fig.~\ref{figure3}f. When further increasing the gas backing pressure to 4~bar, the beam size on the detector increases again (Fig.~\ref{figure3}g). In this case, the focal plane shifts closer to the jet, and a divergent beam is detected. 

\begin{figure}[tb]
 \centering                                          
  \includegraphics[width=8.6cm]{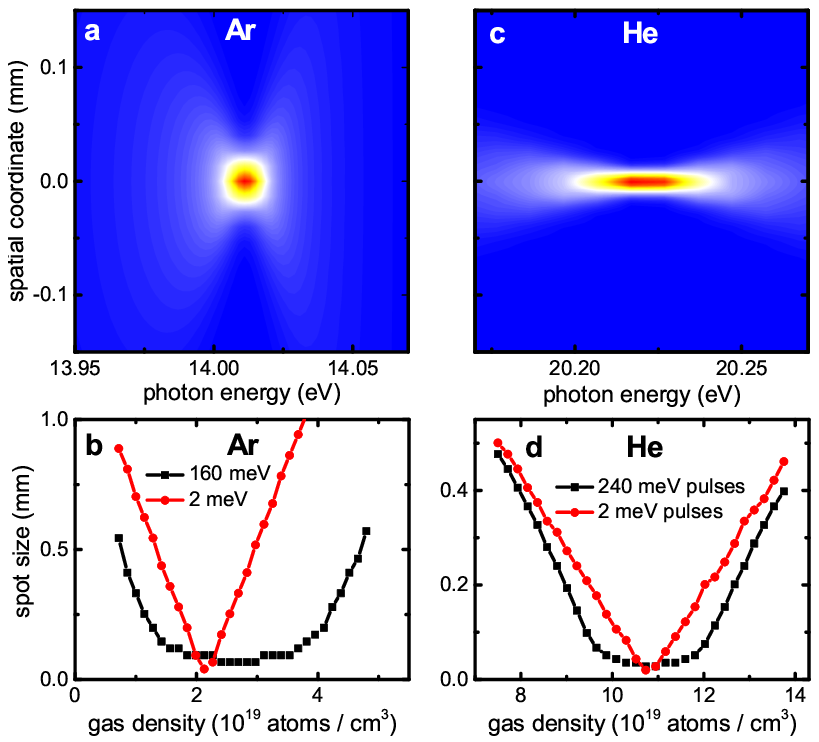}
 \caption{\label{figure4} \textbf{Simulation of the XUV focus.} \textbf{a}, Simulated focus in the vertical direction as a function of the photon energy following propagation of a 9th harmonic XUV pulse at 14.015~eV (1.9~mm FWHM diameter) through an Ar gas jet with a peak density of $2.2\times 10^{19}$~atoms/cm$^3$ (corresponding to a pressure of 0.9~bar at 300~K). Due to chromatic aberration, the XUV spot size depends on the photon energy. \textbf{b}, Spot size as a function of the Ar gas density for XUV pulses with a bandwidth of 160~meV (black curve) and 2~meV (red curve), showing minimal spot sizes of 74 and 40~$\mu$m, respectively \textbf{c}, The chromatic aberration is reduced for photon energies that are further away from the resonance. This is shown for the example of a 13th harmonic XUV pulse at a photon energy of 20.22~eV (2.4~mm FWHM diameter) that propagates through a He gas jet with a peak density of $1.0\times 10^{20}$~atoms/cm$^3$ (corresponding to a pressure of 4.3~bar at 300~K). \textbf{d}, Spot size as a function of the He gas density for XUV pulses with a bandwidth bandwidth of 240~meV (black curve) and 2~meV (red curve), which exhibit minimal spot sizes of 28 and 20~$\mu$m.}
\end{figure}

A minimum beam size of 270~$\mu$m was observed in the experiments with the Ar lens (Fig.~\ref{figure3}h), which is sufficiently small for many applications such as photoion and photoelectron spectroscopy. Some applications such as high-spatial resolution imaging, the investigation of XUV-induced nonlinear processes~\cite{tzallas03, takahashi13, schutte14a, flogel17} and single-shot coherent diffractive imaging~\cite{rupp17} require substantially smaller XUV spot sizes though. The achievable focal spot size is limited by geometric and chromatic aberrations. For ideal focusing conditions, a parabolic profile of the gas density integrated along the XUV beam propagation axis is required. While the gas density profile generated by a cylindrical nozzle is parabolic to a good approximation~\cite{semushin01}, deviations from the parabolic shape lead to geometric aberrations, affecting the focal spot size that can be achieved. Furthermore, a density gradient is present along the gas beam propagation axis that leads to geometric aberration. In the future, the gas density profile may be optimized by tailoring the gas nozzle designs (see e.g.~\cite{semushin01}). 

Assuming an ideal parabolic gas density profile, we have simulated the spot sizes achieved by an Ar lens for a collimated XUV beam at 14.015~eV with a FWHM diameter of 1.9~mm. The XUV spot size at a distance of 90~cm behind the gas lens depends on the photon energy, as shown in Fig.~\ref{figure4}a. This chromatic aberration, which is a direct consequence of the variation of the refractive index within the XUV bandwidth, results in a spot size that is larger than that of a monochromatic XUV beam. Note that this effect is not visible in the experimental data due to the spectral resolution of about 100~meV and the spatial resolution of about 100~$\mu$m. The gas-density dependent spot size at a distance of 90~cm from an Ar lens is plotted in Fig.~\ref{figure4}b, showing a minimum spot size of 74~$\mu$m for an XUV pulse with a bandwidth of 160~meV, which is similar to the bandwidth of the 9th harmonic observed in the experiment (black curve in Fig.~\ref{figure4}c). When reducing the XUV bandwidth to 2~meV, a minimum spot size of 40~$\mu$m is obtained (red curve in Fig.~\ref{figure4}b). The chromatic aberration is reduced when exploiting refraction due to a resonance that is further away from the XUV photon energy, as shown for a He lens with a FWHM diameter of 2.4~mm in Fig.~\ref{figure4}c. A minimum spot size of 28~$\mu$m is obtained in this case using a pulse with a bandwidth of 240~meV (similar to the bandwidth of the 13th harmonic used in the experiment shown in Fig.~3, black curve in Fig.~\ref{figure4}d), and it is reduced to 20~$\mu$m for a pulse with a bandwidth of 2~meV (red curve in Fig.~\ref{figure4}d). Smaller spot sizes can be achieved by further increasing the gas pressure, leading to a shorter focal length of the lens.



When using a refractive lens to focus ultrashort XUV pulses, another important aspect is the XUV pulse duration at the focus. Since HHG and free-electron laser pulses have an intrinsic negative chirp~\cite{schafer97, fruhling09}, a refractive lens, which induces a positive chirp, can lead to compression of the XUV pulses. Assuming a pulse with a duration of 24~fs and a chirp of $-8$~meV/fs, which is in the range of previous measurements of the 13th harmonic~\cite{mauritsson04}, our simulations show compression to 16~fs by a He lens with a peak gas density of $4.9\times 10^{19}$~atoms/cm$^3$ (corresponding to a focal length of 1.9~m). Note that this value is larger than the Fourier-limited pulse duration of 8~fs due to the nonlinear chirp that is introduced by the lens. When increasing the peak gas density to $1.0\times 10^{20}$~atoms/cm$^3$ (corresponding to a focal length of 90~cm), we predict a moderate stretching from 24~fs to 29~fs. Focusing of shorter XUV pulses may be achieved by combining a refractive lens with another focusing element. For example, the development of a multi-component lens consisting of an XUV refractive lens and a Fresnel zone plate was suggested~\cite{wang03,pan16}. It was theoretically shown that these multi-component lenses can be used to focus broadband attosecond pulses to nanometer spot sizes~\cite{pan16}, which may enable the investigation of electronic processes with attosecond temporal and nanometer spatial resolution.

\section*{Summary and Outlook}

\noindent In this paper, we have presented a method to deflect and focus XUV pulses by exploiting the inhomogeneity of a gas jet that is placed in the way of an XUV beam. Our results enable the transfer of concepts based on refractive optics that are widely used in other spectral regions, to the XUV regime, including microscopy, nanofocusing and the compression of ultrashort pulses. XUV lenses based on a gas target have several advantages, including their high transmission, deformability and tunability (by varying the gas composition, the gas pressure and the gas jet geometry). Compared to reflective mirrors that are used to focus XUV pulses, these XUV lenses are immune to damage (since the gas sample is constantly replenished), and preserve the propagation direction of the incoming XUV light, thereby facilitating their use in experimental setups.

While in the current paper we have emphasized the refraction of XUV pulses as a result of the stationary refractive index of the gas medium used, in the future, our method may be extended to study and exploit transient refractive index changes in the XUV regime. Examples are the Kerr effect that induces a nonlinear refractive index in the presence of a strong electric field, as well as refraction due to free electrons in a dense plasma, which may be used to develop a plasma lens for XUV and soft X-ray pulses.

\section*{Methods}

\noindent \textbf{XUV prism experiments.} The XUV prism experiments were performed at an HHG beamline that was previously described in detail~\cite{neidel13, drescher16, galbraith17}. NIR pulses with a duration of 4.5~fs were obtained by spectrally broadening the output of a commercial Ti:sapphire amplifier in a differentially pumped hollow-core fiber that was filled with Ne. Temporal compression of the broadened spectrum was achieved using chirped mirrors. High harmonics were generated by focusing the compressed NIR pulses into a gas cell that was filled with Xe. A 100~nm thick Al filter was used to block the NIR beam, and the XUV pulses were refocused by a toroidal mirror. Angle-resolved XUV spectra were recorded by an XUV spectrometer that consisted of a flat-field grating, a multichannel plate / phosphor screen assembly and a digital camera. 

Control over the XUV refraction was achieved by a pulsed gas jet that was positioned near the XUV focus. The gas jet was generated by a piezoelectric valve with a nozzle diameter of 0.5~mm. A three-dimensional manipulator was used to position the gas jet with respect to the XUV focal spot. The XUV beam crossed the gas beam at a distance of about 100~$\mu$m from the exit of the nozzle, where gas densities up to $10^{20}$~cm$^{-3}$ were achieved. When applying a certain backing pressure, we estimate the peak pressure in the interaction region to be smaller by a factor of about 3 (when assuming a temperature of 300~K). The XUV focal spot size ($\approx 100$~$\mu$m) was small compared to the extension of the gas jet at the laser position ($\approx 1$~mm). The gas density gradient along the gas beam propagation direction is estimated to be much smaller than the density gradient perpendicular to this propagation direction.

\vspace{5mm}

\noindent \textbf{XUV lens experiments.} The XUV lens experiments were performed at a second HHG beamline, where harmonics with a narrow bandwidth are available~\cite{schutte14a}. NIR pulses with a pulse energy of 30~mJ and a duration of 35~fs were generated using a home-built Ti:sapphire amplifier~\cite{gademann11}. High harmonics were generated by focusing the NIR pulses using a spherical mirror with a focal length of 5~m into a 10~cm long gas cell that was filled with Xe. 

In order to focus the XUV beam, a piezoelectric valve with a nozzle diameter of 1~mm was positioned at a distance of 5~m behind the generation cell. Before propagating through the gas jet, the HHG beam was truncated by a slit to a horizontal width of 200~$\mu$m to increase the spectral resolution in the spectrometer located downstream. A 100~nm thick Al filter was used in the He experiment, which was removed for the Ar experiment, because it absorbs the 9th harmonic. An XUV spectrometer consisting of a plane grating, a multichannel plate / phosphor screen assembly and a digital camera was used to spectrally and spatially characterize the XUV pulses.


\vspace{5mm}

\noindent \textbf{Simulations.} In order to simulate the refraction of XUV pulses following propagation through an inhomogeneous gas jet, we calculated the complex refractive index $\tilde{n}$ using the Lorentz-Lorenz formula~\cite{born99} 

\begin{equation}
\frac{\tilde{n}^2-1}{\tilde{n}^2+2} = N(x) \frac{e^2}{3 m_e \epsilon_0} \sum_j \frac{f_j}{\omega_{0j}^2-\omega^2-i\Gamma_j \omega}.
\label{equation1}
\end{equation}

\noindent Here $N(x)$ is the (spatially-dependent) atomic density, $e$ is the electron charge, $m_e$ is the electron mass, $\epsilon_0$ is the vacuum permittivity, $f_j$ is the oscillator strength of the transition $j$, $\omega$ is the angular frequency of the XUV light, and $\omega_{0j}$ and $\Gamma_j$ are the resonant frequencies and widths, respectively. The real part of the refractive index, $n$, and the absorption index, $\beta$, were extracted according to the following equations:

\begin{eqnarray}
n = \sqrt{0.5\times (|\tilde{n}|^2+Re\{\tilde{n}^2\})}, \\
\beta = \sqrt{0.5\times (|\tilde{n}|^2-Im\{\tilde{n}^2\})}.
\end{eqnarray}

We used the eikonal approximation (assuming light propagation along straight lines)~\cite{born99} to calculate the phase and the amplitude of the XUV field following propagation through a gas jet with thickness $L$:

\begin{equation}
A(x,z=0) = A_0(x) e^{\frac{\omega L}{c} (i(n-1)-\beta)}.
\end{equation}

\noindent Here $A_0$ is the amplitude of the incoming XUV field. We restricted our calculations of the complex refraction to the $xz$ plane, taking into account only the $x$ dependence of the XUV beam profile. Since the deflection angles observed in the present experiments are small (for example, the maximum deflection angle observed in Fig.~\ref{figure2}a is about 3.5~mrad), we used the Kirchhoff diffraction formula in the small-angle approximation~\cite{born99} to calculate the XUV amplitude at a screen located at a distance $S$ from the jet. The angle-dependent XUV amplitudes were calculated by the following formula:

\begin{equation}
\tilde{A}(\theta,z=S) \propto \int A(x) e^{i\frac{w}{c}\left(x\sin\theta+\frac{x^2}{2S}\right)} dx. 
\end{equation}

In the simulations the temporal envelope of the XUV pulse is calculated by taking the square of the Fourier transform along the spectral axis in Fig.~3 and averaging over the spatial coordinate. As mentioned in the main text, the incoming pulses are assumed to be chirped by about $-8$~meV/fs resulting in an XUV pulse with a duration of 24~fs. Depending on the applied gas density, the pulses in the focus are either compressed or acquire a positive chirp.

\section*{References}

\bibliography{Bibliography}

\section*{Acknowledgements}

This project has received funding from the European Union's Horizon 2020 research and innovation programme under the Marie-Sklodowska-Curie grant agreement no. 641789 MEDEA.

\section*{Author contributions}

L.D. and B.S. performed the prism experiments, and B.S. performed the lens experiments. O.K. carried out the simulations. All authors discussed the results and contributed to writing the manuscript.

\section*{Additional information}

\section*{Competing financial interests}

The authors declare no competing financial interests.

\end{document}